# Use of Robust DOB/CDOB Compensation to Improve Autonomous Vehicle Path Following Performance in the Presence of Model Uncertainty, CAN Bus Delays and External Disturbances

Haoan Wang, Levent Guvenc

Automated Driving Lab, Ohio State University

## Abstract

Autonomous vehicle technology has been developing rapidly in recent years. Vehicle parametric uncertainty in the vehicle model, variable time delays in the CAN bus based sensor and actuator command interfaces, changes in vehicle sped, sensitivity to external disturbances like side wind and changes in road friction coefficient are factors that affect autonomous driving systems like they have affected ADAS and active safety systems in the past. This paper presents a robust control architecture for automated driving systems for handling the abovementioned problems. A path tracking control system is chosen as the proof-of-concept demonstration application in this paper. A disturbance observer (DOB) is embedded within the steering to path error automated driving loop to handle uncertain parameters such as vehicle mass, vehicle velocities and road friction coefficient and to reject yaw moment disturbances . The compensation of vehicle model with the embedded disturbance observer forces it to behave like its nominal model within the bandwidth of the disturbance observer. A parameter space approach based steering controller is then used to optimize performance. The proposed method demonstrates good disturbance rejection and achieves stability robustness. The variable time delay from the "steer-by-wire" system in an actual vehicle can also lead to stability issues since it adds large negative phase angle to the plant frequency response and tends to destabilize it. A communication disturbance observer (CDOB) based time delay compensation approach that does not require exact knowledge of this time delay is embedded into the steering actuation loop to handle this problem. Stability analysis of both DOB and CDOB compensation system are presented in this paper. Extensive model-in-the-loop simulations were performed to test the designed disturbance observer and CDOB systems and show reduced path following errors in the presence of uncertainty, disturbances and time delay. A validated model of our 2017 Ford Fusion Hybrid research autonomous vehicle is used in the simulation analyses. Simulation results verify the performance enhancement of the vehicle path following control with proposed DOB and CDOB structure. A HiL simulator that uses a validated CarSim model with sensors and traffic will be used later to verify the real time capability of our approach.

## I. Introduction

With the rapid development of autonomous vehicles, automatic steering technique plays an important role in autonomous research area. Many different steering control methods have been proposed in the literature. A path following algorithm named Circular Look Ahead (CLA) steering control was proposed in [1] which can control a car to precisely follow a path even on a curvy road. The waypoint tracking method of autonomous navigation is presented in [2] using the Point to Point algorithm with position and heading measurements from GPS receivers. Model predictive control based vehicle front wheel steering is applied to track the collision free path in [3] and has the capability to deal with a wide variety of process control constraints systematically. However, regular controllers are usually designed without considering external disturbances and model uncertainty in mind, which may lead to performance degradation in path tracking. To solve such problem, a disturbance observer (DOB) is added into the control system to achieve insensitivity to modeling error and disturbance rejection. The disturbance observer was firstly proposed by Ohnishi [4] and further developed by Umeno and Hori [5]. Later, DOB has been applied in mechatronic applications in the literature. In [6], robustness of disturbance observer is added to the model of electrohydraulic system considering the case in which the plant has large parametric variation. A new active front steering controller design for electric vehicle stability using disturbance observer was proposed in [7].

Time delay is another significant issue which generally exists in the network-based control system. With the occurrence of time delay, large negative phase angles are added to the frequency response of vehicle plant which may lead to instability of the system. The Smith predictor has been widely used for a long time and extended for different cases such as [8-9]. Smith predictor has the advantage of easy implementation and simplicity in understanding. However, time delay model and model accuracy in the knowledge of time delay are required to ensure no degradation of compensation performance. Communication disturbance observer was proposed as another time delay compensation approach. This method was firstly applied in the bilateral teleoperation systems [10] and has been extended to robust time delayed control system in [11-12]. The communication disturbance observer can be implemented to a wider range of applications since the accuracy of time delay is not necessary and also can be used for plants with variable time delay.

Motivated by the limitations of single DOB and CDOB compensated system. [13] proposed a double disturbance observer (DDOB) structure in the wireless motion control system design, which embedded both DOB and CDOB in one control system. The proposed approach effectively realized time delay compensation and external disturbance rejection simultaneously.

Although DOB and CDOB have been applied to many different applications in the literature, there are few DOB and CDOB applications in autonomous vehicle system which will be a potential area of progress. Furthermore, DOB, CDOB and DDOB compensated structure investigated in this paper was applied in the autonomous vehicle path following control system separately as a new topic in the field of automated vehicle. Uncertain parameters including vehicle mass, vehicle velocities and road friction coefficient and disturbances like road curvatures are firstly focused on. A disturbance observer (DOB) is embedded within the steering to path error automated driving loop to reject disturbances and handle model uncertainty. Then, time delay was taken into account and CDOB was embedded into the steering actuation loop to handle the problem. Robustness of stability of both structures is analyzed and validated. In order to deal with time delay and external disturbances simultaneously, DDOB compensated structure was used. Simulation results show that DDOB works better than DOB or CDOB compensated systems and all three compensated systems demonstrate good path following performance compared with PD feedback control system.

The rest of this paper is organized as follows. The vehicle steering model and vehicle parameters are presented in Section II. Disturbance observer and communication disturbance observer and are introduced in Section III and Section IV respectively. In Section V, robust PD controller and Q filter are designed. Also, robust stability analysis of both DOB and CDOB design are demonstrated. Section VI proposed double disturbance observer and Section VII shows autonomous vehicle path following simulation results using DOB compensation system, CDOB compensation system and results comparison between DDOB and CDOB. The paper ends with conclusion and recommendations for future work in Section VII.

## II. Vehicle Model

By combining the two front wheels together and two rear wheels together of a four wheel car, a single track vehicle model is formed as shown in Figure 1 to model the steering dynamics. The parameters of the vehicle model are given in Table 1.The state space model can be described as:

$$\begin{bmatrix} \dot{\beta} \\ \dot{r} \\ \Delta\dot{\psi} \\ \dot{y} \end{bmatrix} = \begin{bmatrix} a11 & a12 & 0 & 0 \\ a21 & a22 & 0 & 0 \\ 0 & 1 & 0 & 0 \\ V & l_s & V & 0 \end{bmatrix} \begin{bmatrix} \beta \\ r \\ \Delta\psi \\ y \end{bmatrix} + \begin{bmatrix} b11 \\ b21 \\ 0 \\ 0 \end{bmatrix} \delta_f + \begin{bmatrix} 0 \\ 0 \\ -\mathrm{V} \\ 0 \end{bmatrix} \rho_{ref}$$

(1)

where

a11=$-(c_r+c_f)/\tilde{m}V$, a12=-1+$(c_r l_r - c_f l_f)/\tilde{m}V^2$ (2)

a21=$(C_r l_r - c_f l_f)/J$, a22=$-(c_r l_r^2 + c_f l_f^2)/JV^2$

b11=$c_f/\tilde{m}V$, b12=$c_f lf/J$

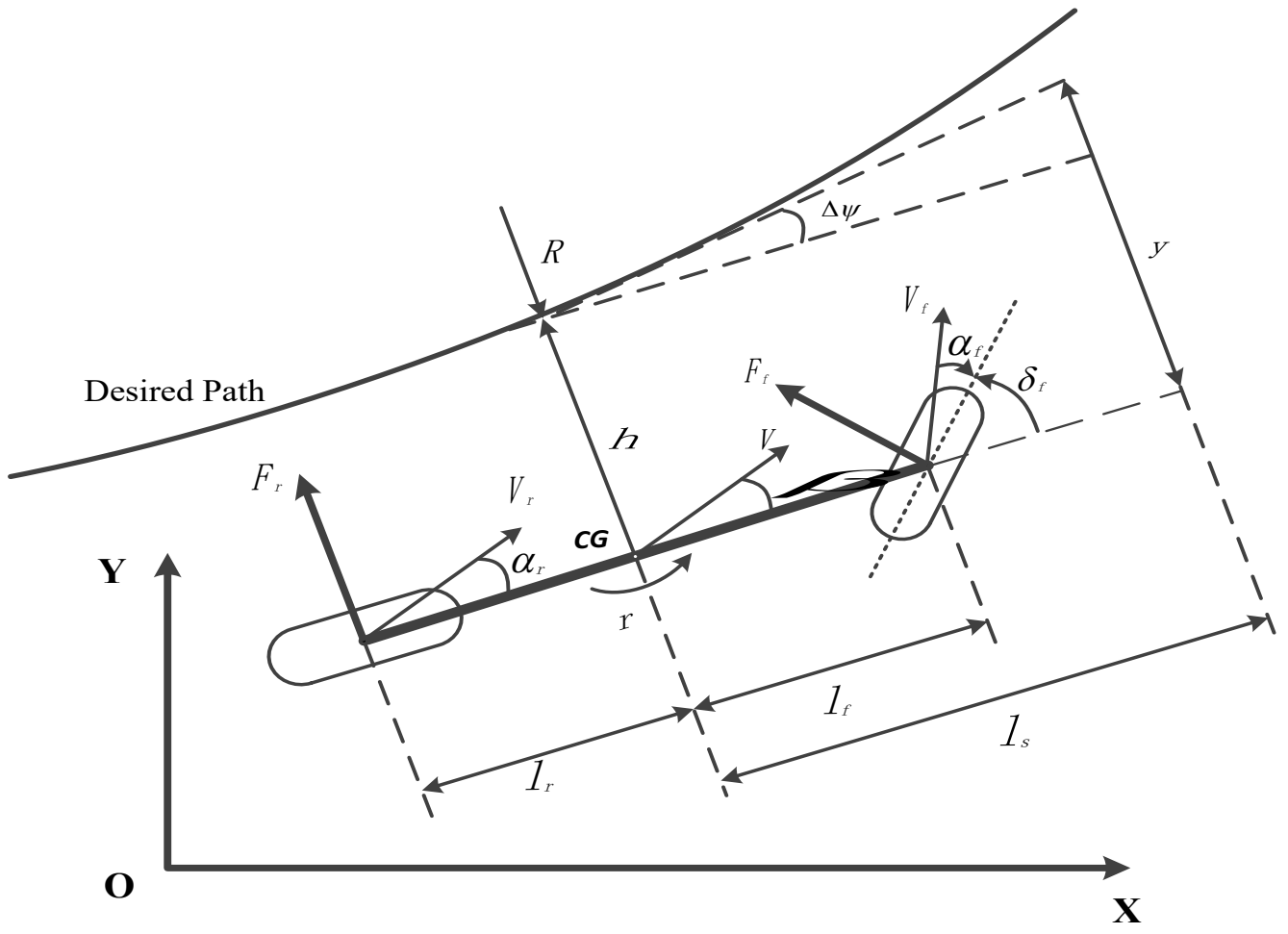

Figure 1.Diagram of the vehicle model

The standard form of vehicle steering dynamics can be written as (3) according to (1):

$\dot{x} = \mathrm{Ax} + \mathrm{Bu}$ (3)

The transfer function from front wheel steering angle $\delta_f$ to the lateral deviation y is represented by equation (4). Note that front wheel steered vehicle is considered in this paper so that δr=0.

$$\frac{y}{\delta_f} = G_{\delta_f} = [0\ 0\ 0\ 1](sI - A)^{-1} \begin{bmatrix} b11 \\ b21 \\ 0 \\ 0 \end{bmatrix} \quad (4)$$

The curvature $\rho_{ref}$ of the desired path is taken as an external disturbance. The transfer function from the road curvature $\rho_{ref}$ to the lateral deviation from the desired path can be represented as:

$$\frac{y}{\rho_{ref}} = G_{\rho_{ref}} = [0\ 0\ 0\ 1](sI - A)^{-1} \begin{bmatrix} 0 \\ 0 \\ -V \\ 0 \end{bmatrix} \qquad (5)$$

The vehicle velocity $V$, vehicle virtual mass m and road friction coefficient μ are regarded as uncertainty parameters with nominal parameter values of $V_n = 5km/h$, $\mu_n = 1$ and $m_n = 2{,}000\ kg$. The operating ranges used were V∈ $[4,7]km/h$ , μ ∈ [0.4,1] and mass m ∈ [1600,2000] kg from no load to full load. The virtual mass $\tilde{m} = \frac{m}{\mu}$ is then within the range of $\tilde{m} = \frac{m}{\mu} \in [1600{,}5000]$ kg. The uncertainty parameters are illustrated in the uncertainty box shown in Figure 2. Four vertices labeled by a, b, c, d in the uncertainty box are used to evaluate the performance and robustness of the disturbance observer compensated system.

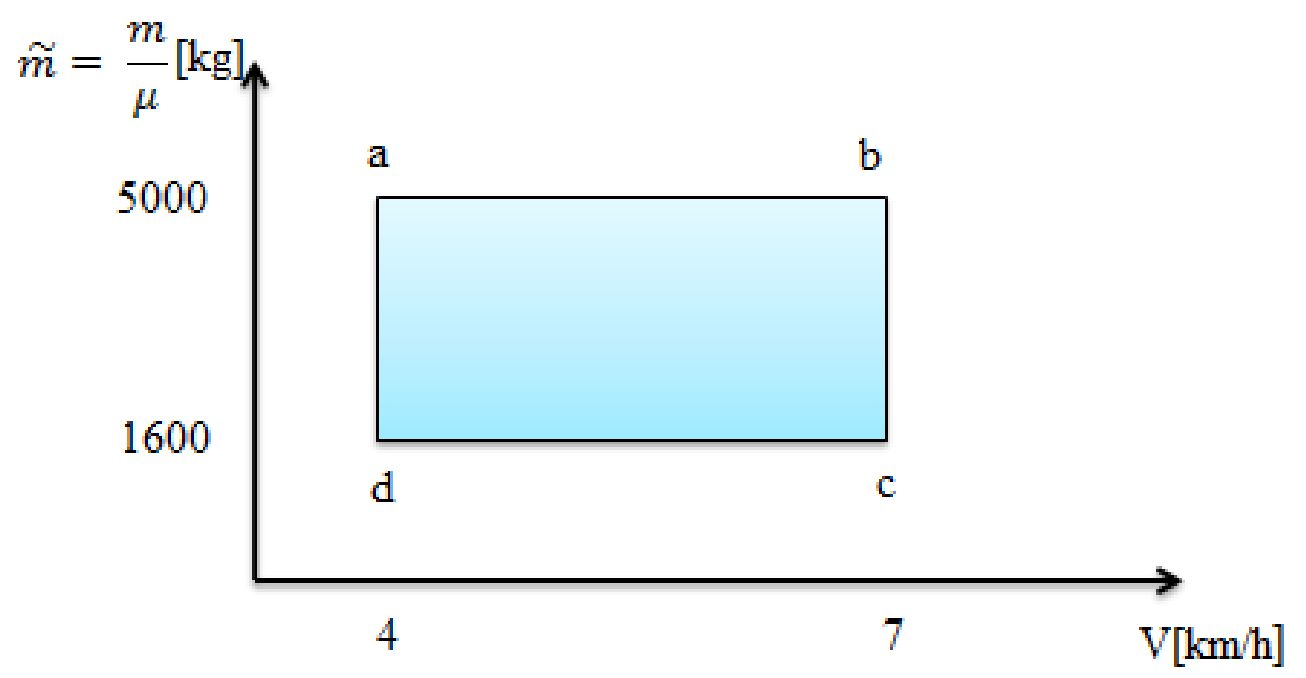

Figure.2 Parametric Uncertainty Box

Table I. Parameters of the vehicle model

| | |
|---|---|
| *β* | vehicle side slip angle [rad] |
| subscript *f* | front tires |
| *V* | vehicle velocity [m/s] |
| $\delta_f$ | front wheel steering angle [rad] |
| *J* | yaw moment of inertia [3728 kgm$^2$] |
| $C_r$ | rear cornering stiffness [50,000 N/rad] |
| $l_f$ | distance from CG to front axle [1.3008 m] |
| $l_r$ | distance from CG to rear axle [1.5453 m] |
| $\rho_{ref}=1/R$ | curvature of path [1/m] |
| *r* | vehicle yaw rate [rad/s] |
| subscript *r* | rear tires |
| $\Delta\psi$ | yaw orientation error with respect to path [rad] |
| *y* | lateral deviation [m] |
| $C_f$ | front cornering stiffness [195,000 N/rad] |
| $m$ | vehicle mass [2,000 kg] |

## III. Disturbance Observer

The block diagram of the closed-loop control system with disturbance observer compensation is depicted in Figure 3. In the block diagram, robust PD feedback controller is used as a baseline controller which is designed based on the nominal model of the vehicle. *Q* is the low pass filter to be selected and its bandwidth determines the bandwidth of model regulation and disturbance rejection. System plant *G* is formulated by taking both model uncertainty $\Delta_m$ and external disturbance *d* into account. The vehicle input - output relation becomes

$$y = Gu + d = (G_n(1 + \Delta m))u + d \qquad (6)$$

where $G_n$ is the desired model of plant and $G$ represents the actual plant. The goal in disturbance observer design is to obtain

$$y = G_n u_{new} \qquad (7)$$

as the input-output relation in the presence of model uncertainty $\Delta_m$ and external disturbance *d*. $u_{new}$ is regarded as a new steering input which is derived as follows. By considering model uncertainty and external disturbance as an extended disturbance *e*, equation (6) can be rewritten as (8)

$$y = (G_n(1 + \Delta m))u + d = G_n u + e \qquad (8)$$

Combining equation (7) with equation (8), the new control input $u_{new}$ is represented as

$$u_{new} = u + \frac{e}{G_n} \qquad (9)$$

and

$$u = u_{new} - \frac{e}{G_n} = u_{new} - \frac{y}{G_n} + u \qquad (10)$$

In order to limit the compensation to a low frequency range to avoid stability robustness problem at high frequency, the feedback signals in (10) are multiplied by the low pass filter Q and implementation equation becomes

$$\mathrm{u} = \mathrm{u}_{new} - \frac{Q}{G_n} y + Qu \qquad (11)$$

Based on the block diagram, the model regulation and disturbance rejection transfer function can be derived as equations (12) (13). It can be seen that Q should be a unity gain low pass filter to make sure as Q→ 1, $\frac{y}{u_{new}} \to G_n$ for model regulation and $\frac{y}{d} \to 0$ to achieve disturbance rejection.

$$\frac{y}{\mathrm{u}_{new}} = \frac{G_n G}{GQ + G_n(1-Q)} \qquad (12)$$

$$\frac{y}{d} = \frac{G_n(1-Q)}{GQ + G_n(1-Q)} \qquad (13)$$

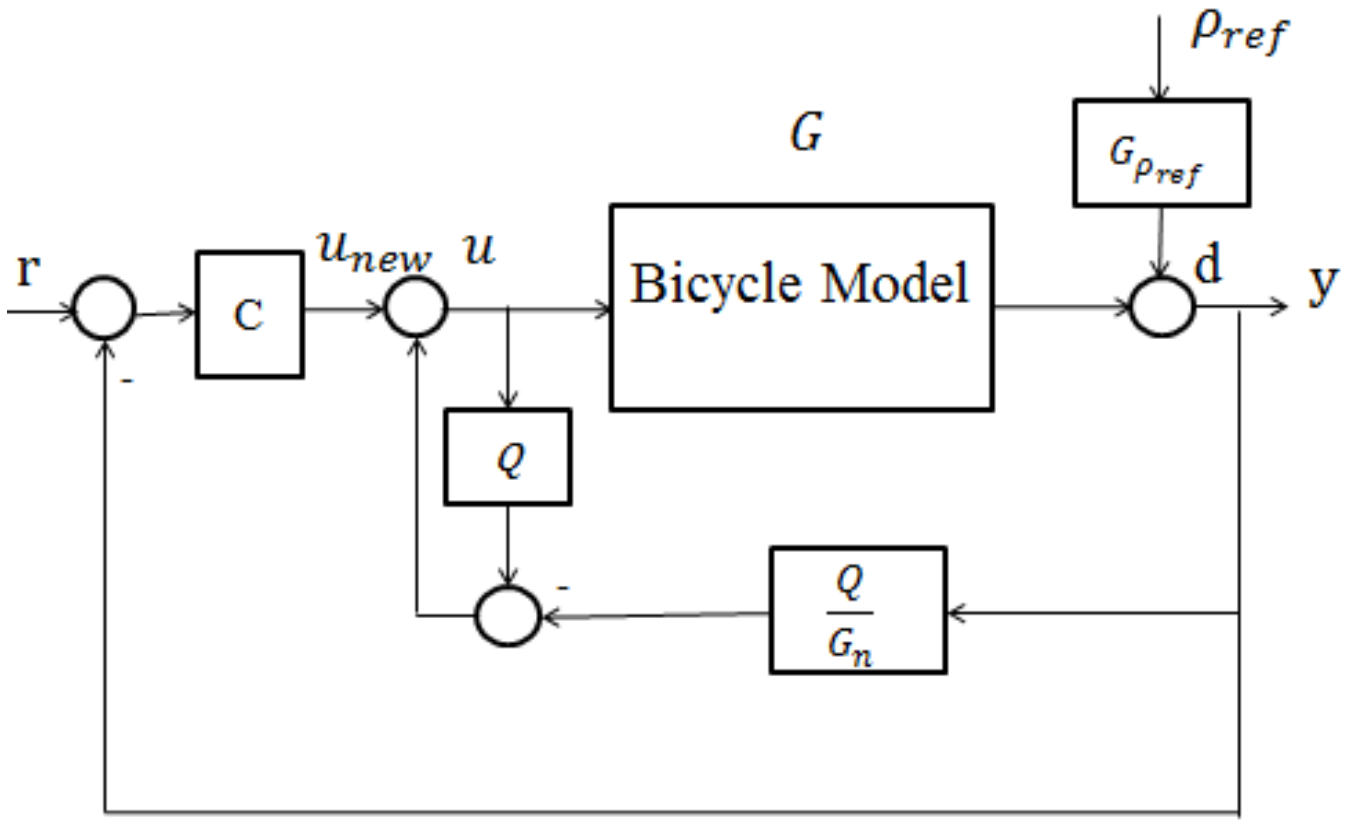

Figure 3. Disturbance observer compensated control system

## IV. Communication Disturbance Observer

Although disturbance observer shows good performance in model regulation and disturbance rejection, performance will degrade when there exists time delay in the system. Communication disturbance observer is applied to compensate the time delay. For CDOB design, time delay is considered as a disturbance $d$ that is acting on the system as illustrated in Figure 4 and the aim is to obtain disturbance estimation $\hat{d}$. From Figure 4, we can get equation (14) and it can be rewritten as (15). Then, the estimated disturbance $\hat{d}$ is obtained by multiplying $d$ with $Q$ to ensure causality as shown in equation (16).

$$y = G_n(u - d) \qquad (14)$$

$$\mathrm{d} = \mathrm{u} - G_n^{-1} y \qquad (15)$$

$$\hat{d} = \mathrm{Q}\,(\mathrm{u} - G_n^{-1} y) \qquad (16)$$

According to network disturbance concept as depicted in Figure 5, $\hat{d}$ can be also expressed as equation (17)

$$\hat{d} = u - ue^{-T} \qquad (17)$$

where $u$ is system input and $T$ is time delay.

In this way, the estimated disturbance $\hat{d}$ is used to compensate the time delay effect in the feedback signal. Figure 6 shows the structure of the communication disturbance observer compensated control system. There is a 0.08 sec time delay between actual steering wheel input and desired steering wheel input, which is compensated by the proposed CDOB. It is seen that there are two blocks in the structure: the left block is time delay compensation and the right block is network disturbance estimation.

Therefore, the closed loop transfer function of the system is written as (18):

$$\frac{y}{r} = \frac{CG_n(s)e^{-Ts}}{1 + CG_nQ + CG_n(1-Q)e^{-T}} \qquad (18)$$

The $Q$ filter is usually chosen as a low pass filter due to the fact that reference operates in low frequency. From equation (18), we can see that it is ideal to make $Q$=1 in low frequency so that the denominator of the transfer function will have no time delay elements.

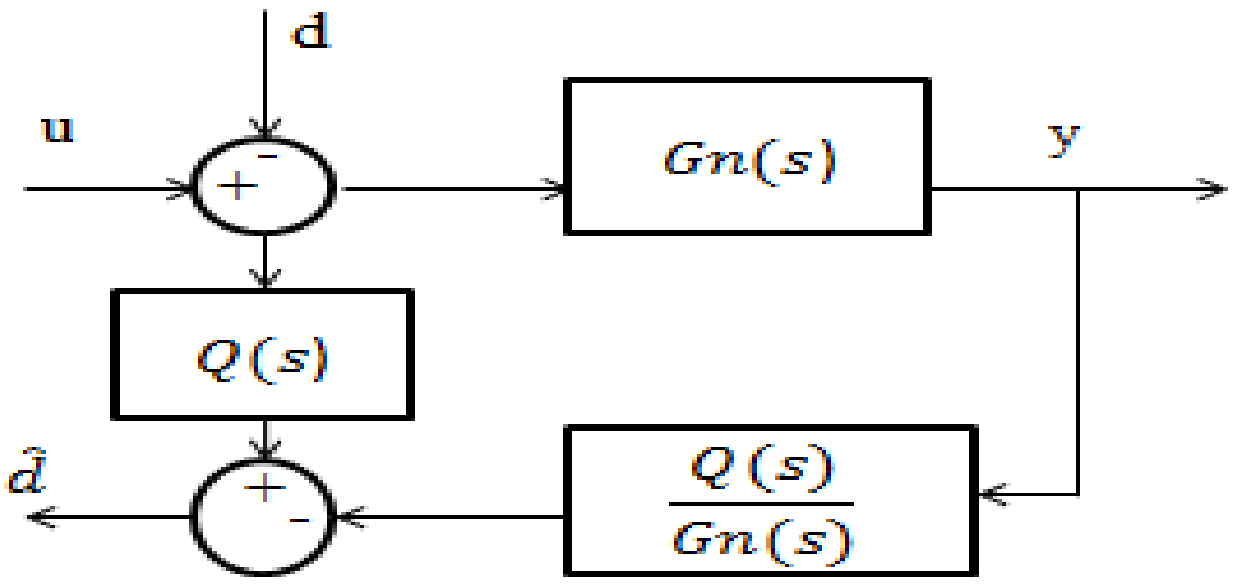

Figure 4. Classic disturbance observer

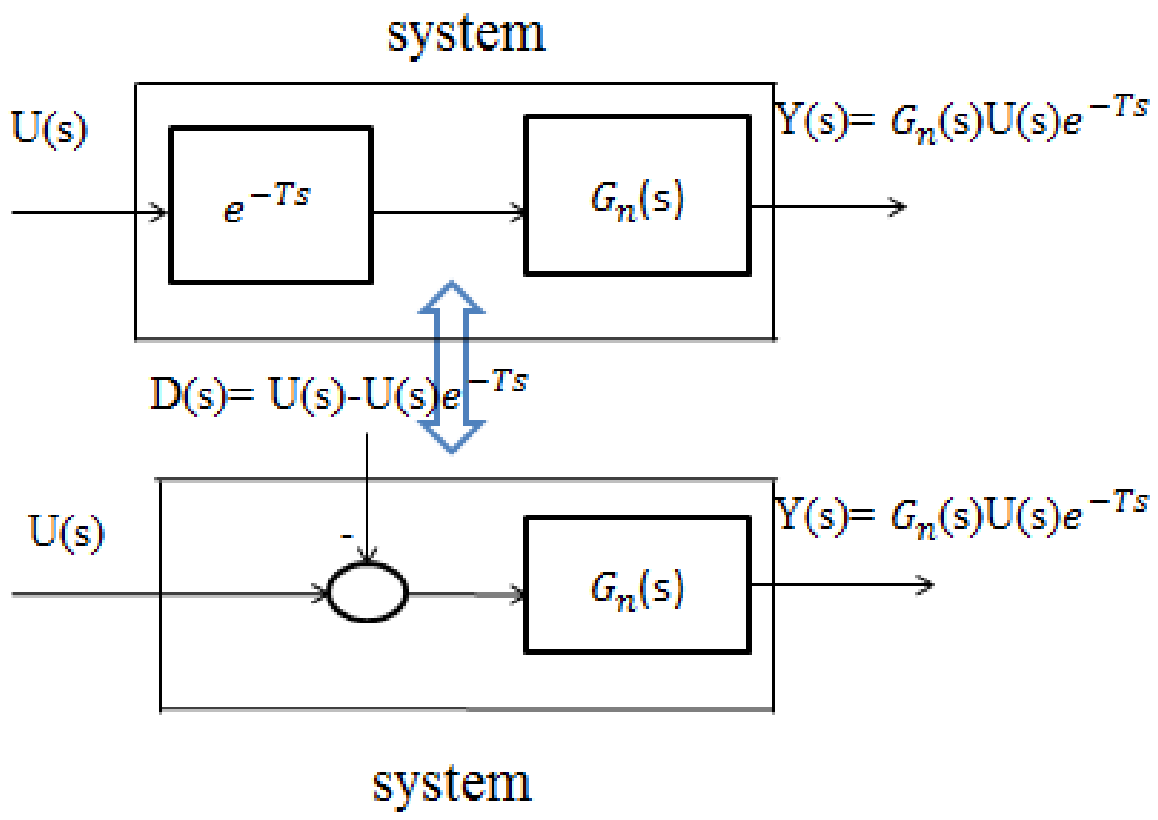

Figure 5. Concept of network disturbance

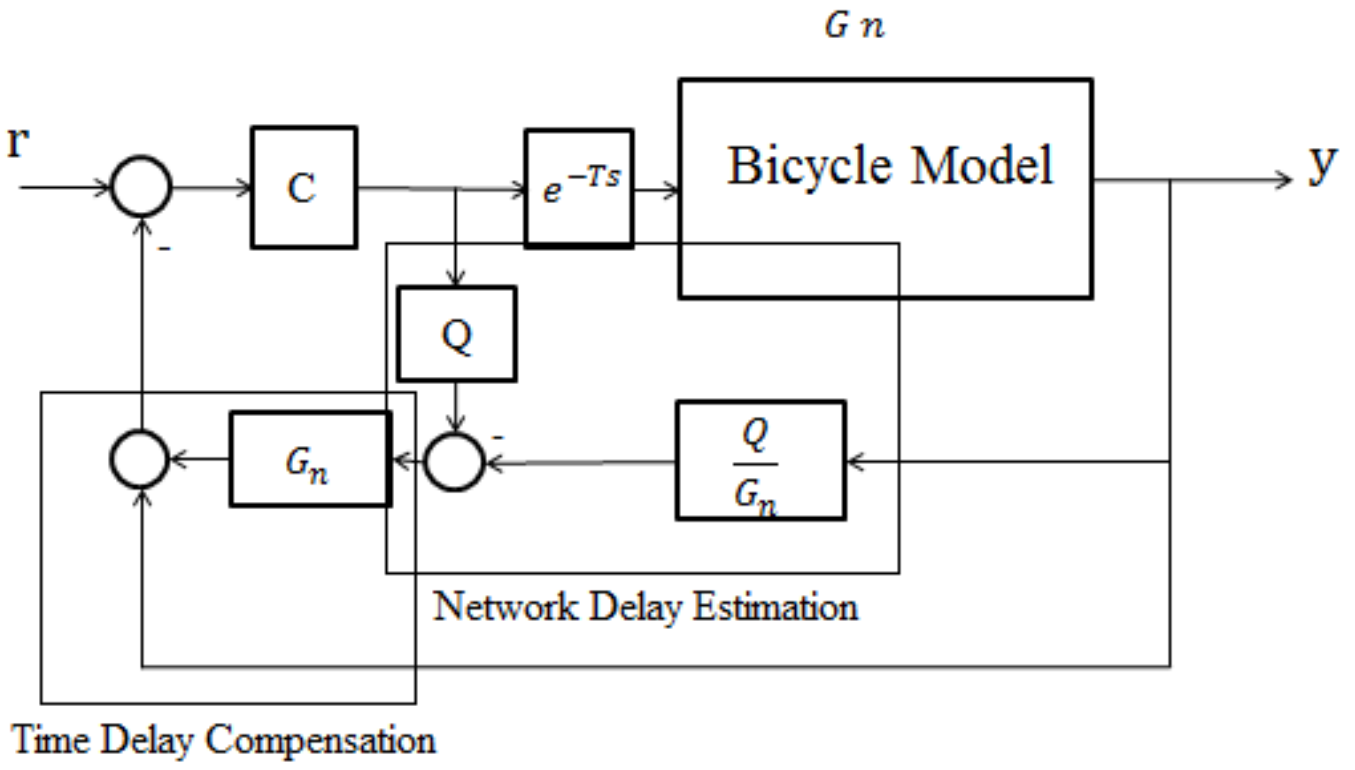

Figure 6. Communication disturbance observer compensated control system

# VI. Design Analysis

## A. Robust PD controller design

In the proposed robust control system, a parameter space approach based PD controller is designed. The details of parameter space method can be found in [14-16]. Robust PD controller is designed based on the nominal plant $G_n$. Using the parameter space method, D-stability boundaries are depicted in Figure 7, where settling time constraint σ is set to be 0.3, damping constraint θ is 135° and bandwidth constraint R is assigned as 1.3 rad/sec. The overall solution region which satisfies the stability requirements are calculated and plotted as illustrated in Figure 8. In Figure 8, $K_p$ and $K_d$ are two free design parameters and we select $K_p$=1.0596, $K_d$=0.939.

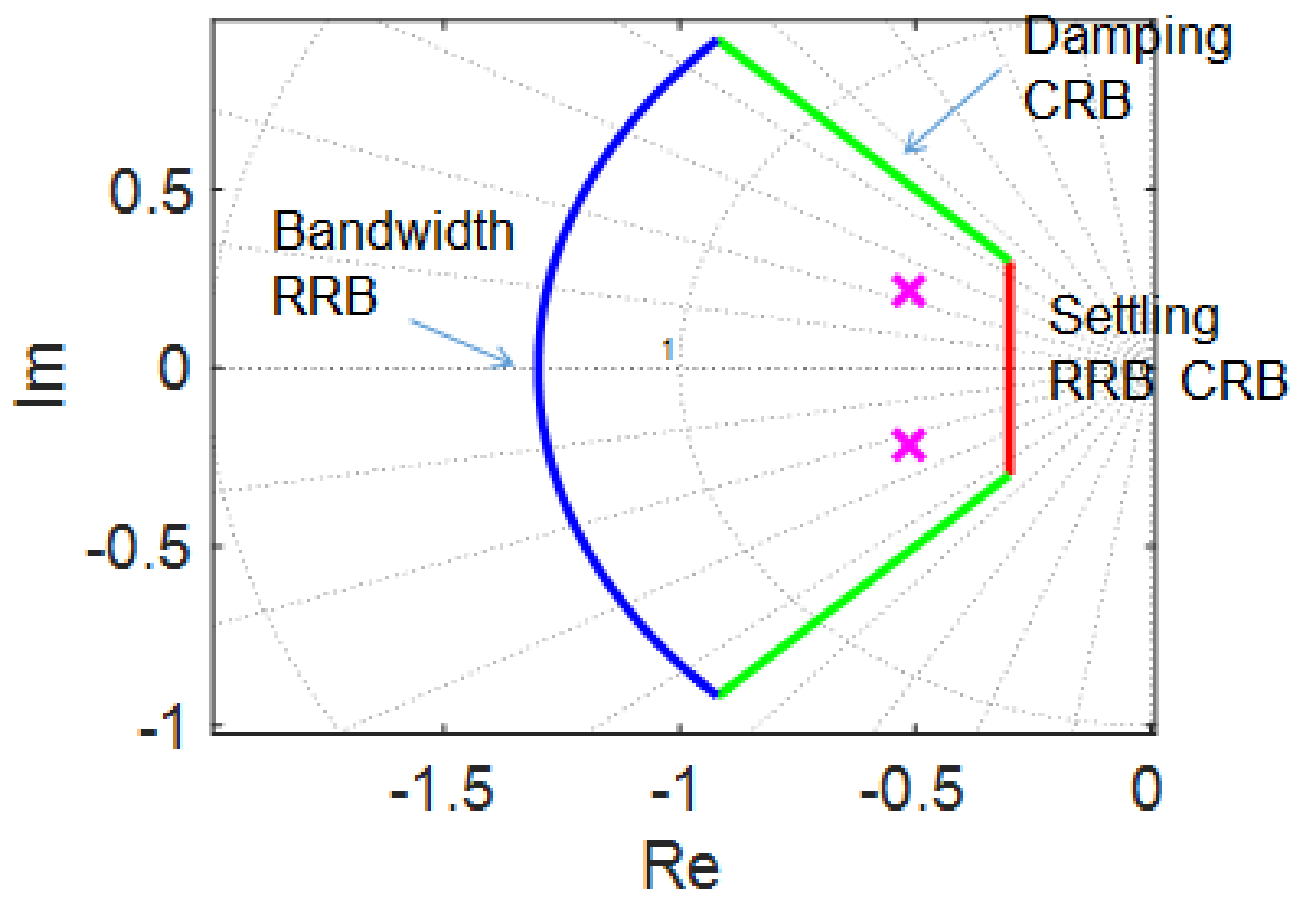

Figure 7. D-stability boundary

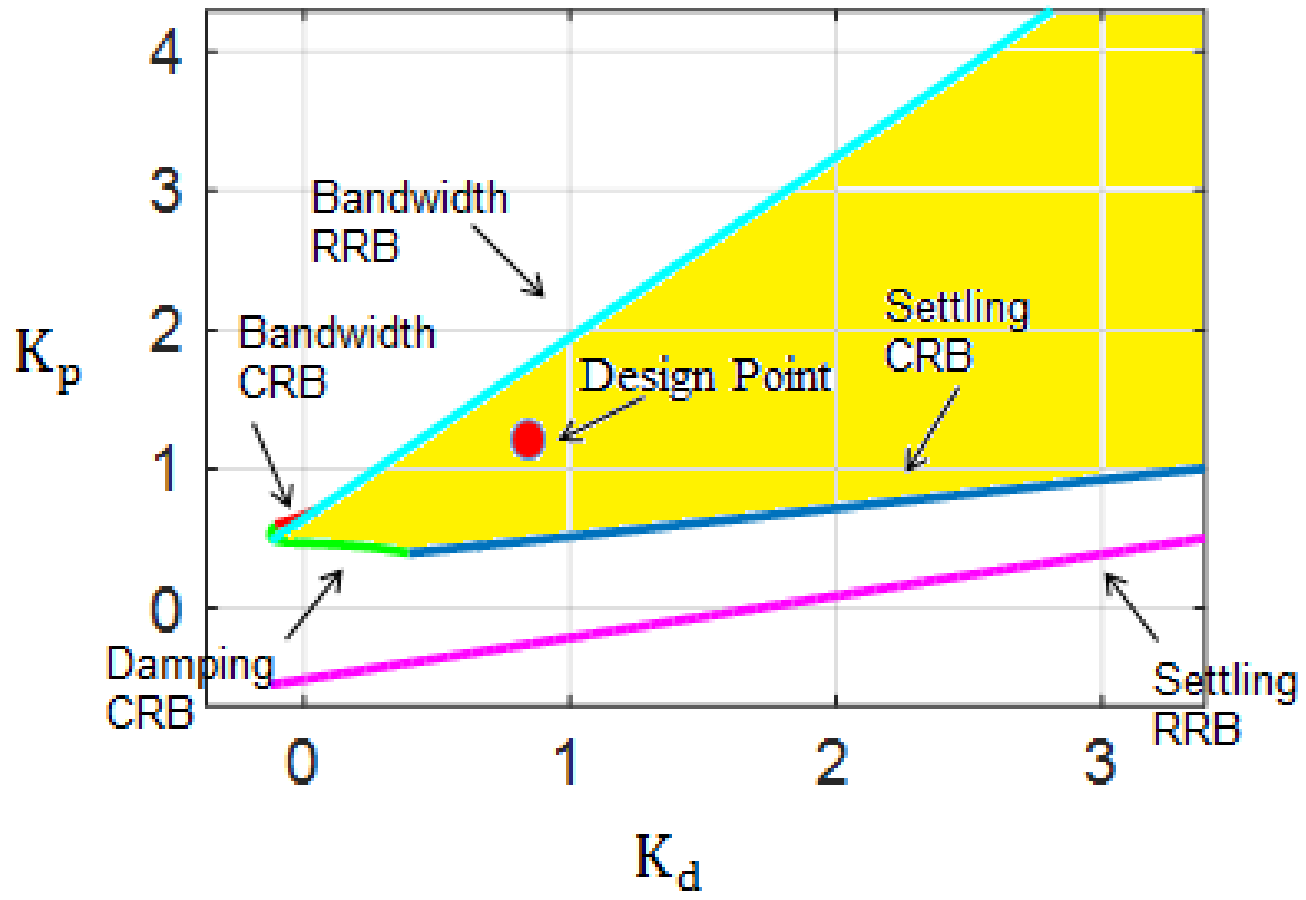

Figure 8. D-stability solution region

## B. Q filter design and verification of robust stability

$Q$ filter is designed to be a low pass filter as discussed before for model regulation, disturbance rejection and time delay compensation. For appropriate orders of the $Q$ filter, since the relative degree of low pass filter $Q$ is chosen to be at least equal to the relative degree of $G_n$ for causality of $Q/G_n$. The vehicle path following transfer function model $G_n$ obtained from equation (4) is calculated as equation (19). Therefore, a second order filter $Q$ is designed as defined in equation (20). For the cutoff frequency of $Q$ filter, it should be appropriately selected in order to make ascertain the stability robustness of the system.

$$\mathrm{Gn(s)} = \frac{227.6\,s^2 + 8.479*10^4\,s + 3.627*10^4}{s^4 + 459.2s^3 + 3.329e04\,s^2} \tag{19}$$

$$\mathrm{Q(s)} = \frac{1}{(\tau s+1)^2} \tag{20}$$

where $\tau=1/\omega_c$.

### B.1 DOB compensation system robust stability analysis

We have obtained that $Q$ must go to unity for model regulation and disturbance rejection. According to the characteristic equation (21), equation (22) is derived since $Q \to 1, G_n(1-Q) \to 0$.

$$G_n(1-Q) + G_n(1+\Delta_m)Q = 0 \tag{21}$$

$$G_n(1+\Delta_m Q) = 0 \tag{22}$$

Based on the small gain theory [17], the sufficient condition for robust stability can be written as equation (23). Combining variations covering all vertices from uncertainty box in Figure 2, real parametric variation of vehicle mass $m$, vehicle velocity $V$ and road friction coefficient $\mu$ are converted to an approximate unstructured multiplicative uncertainty $\Delta_m$. Figure 9 illustrates the satisfaction of disturbance observer design requirement when the cutoff frequency $\omega_c$ of $Q$ is 5 rad/s.

$$|Q| < \left|\frac{1}{\Delta_m}\right|, \forall\omega \qquad (23)$$

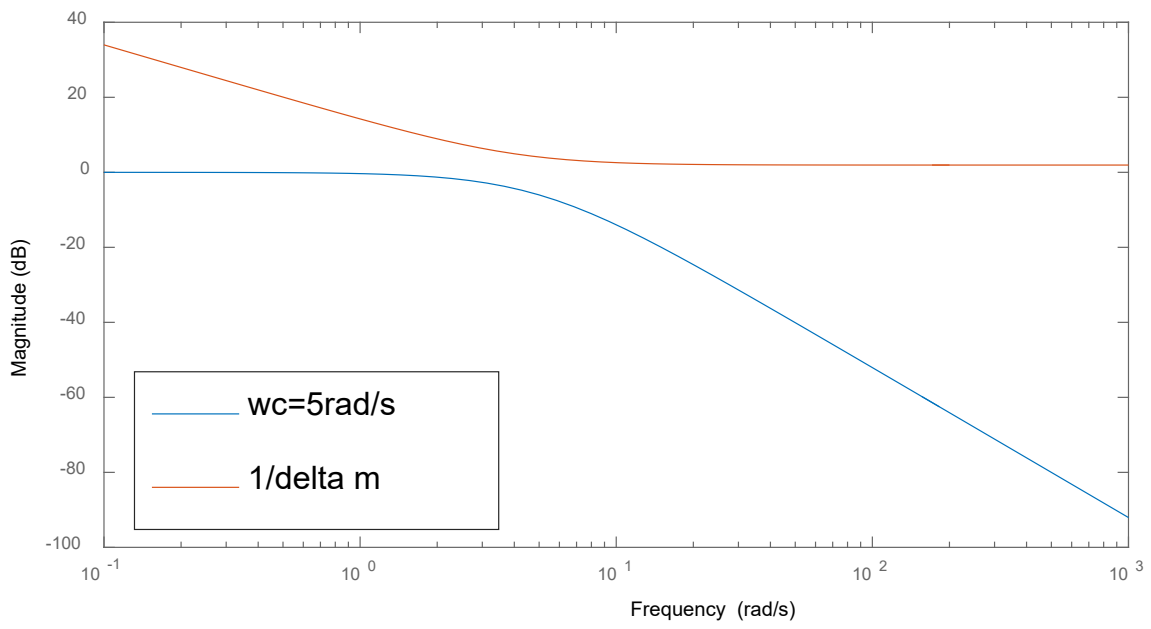

Figure 9. Magnitude of Q and $\frac{1}{\Delta_m}$ for stability of robustness

The frequency responses of four corners of the uncertainty box are also studied to illustrate the robustness of DOB compensated system. PD feedback controller was applied to both systems with and without DOB compensation, the input-output behavior $|y/r|$ are shown below. It can be seen that at low frequency there are larger variations in figure 10 as the operating point is varied than in second figure. In figure 11, the frequency response magnitudes are close to each other at low frequency.

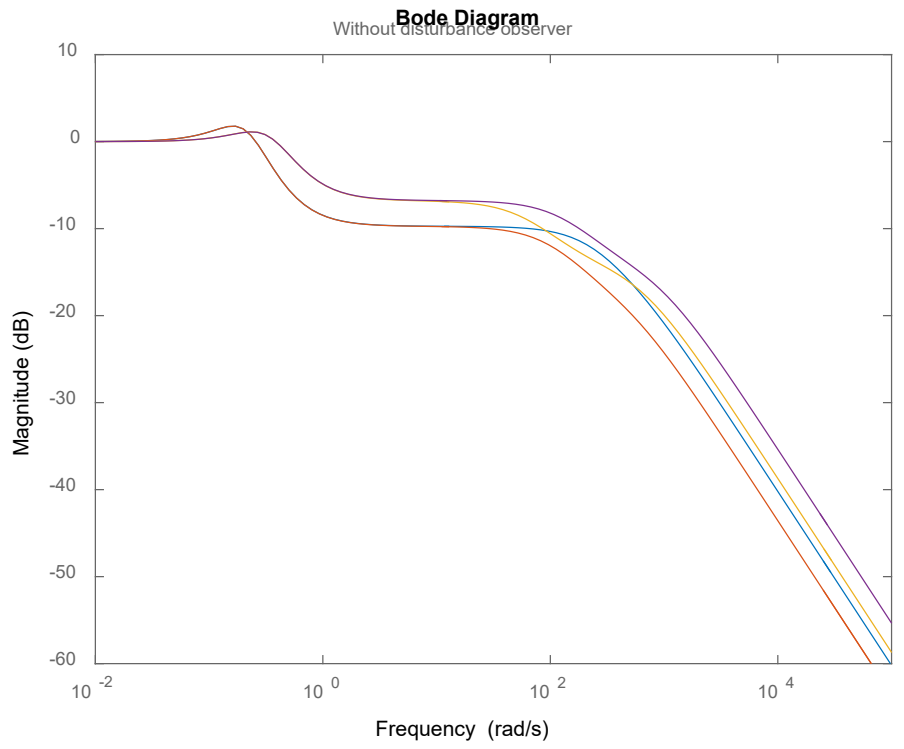

Figure 10 . $|y/r|$ for the four vertices of uncertainty box without disturbance observer

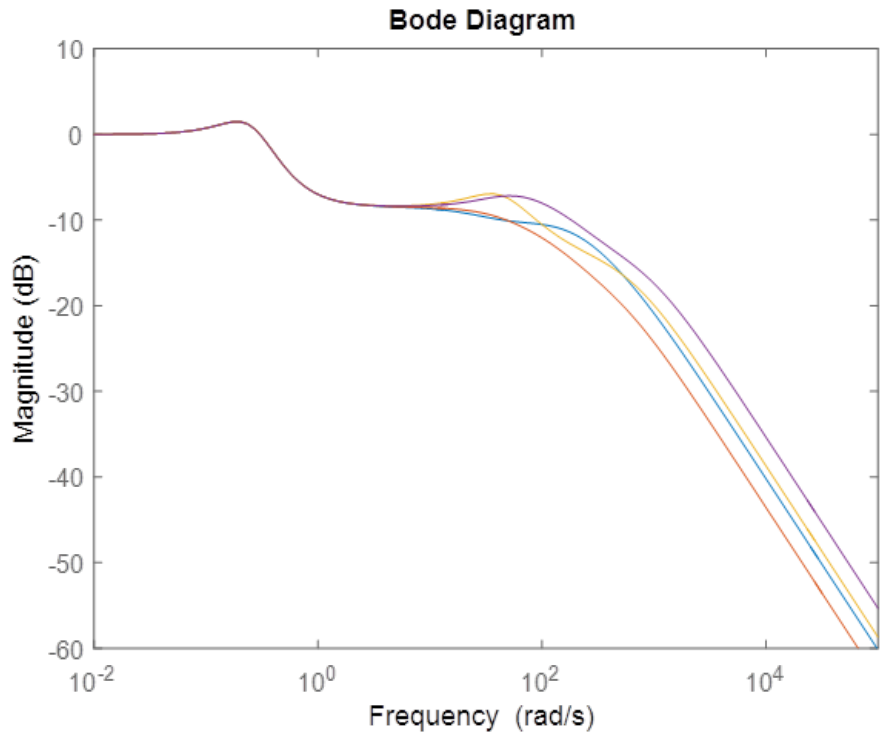

Figure 11. $|y/r|$ for the four vertices of uncertainty box with disturbance observer

### B.2 CDOB compensation system robust stability analysis

According to the Nyquist stability criterion, robust stability of uncertain system can be guaranteed if $L$ does not encircle point (-1, 0), which can be expressed as equation (24):

$$|\Delta_m(j\omega)L_n(j\omega)| < |1 + L_n(j\omega)|, \forall\omega \qquad (24)$$

or equivalently,

$$\left|\frac{\Delta_m(j\omega)L_n(j\omega)}{1+L_n(j\omega)}\right| < 1, \forall\omega \leftrightarrow \left|\frac{L_n(j\omega)}{1+L_n(j\omega)}\right| < \left|\frac{1}{\Delta_m(j\omega)}\right|, \forall\omega \qquad (25)$$

where $L_n$ is represented as in equation (26) in this system, which is the nominal loop transfer function.

$$L_n = \frac{C(1-Q)G_n e^{-Ts}}{1+CG_nQ} \qquad (26)$$

Consider time delay $e^{-Ts}$ as the source of unmodeled dynamics, the model uncertainty $\Delta_m$ is then given by equation (27):

$$\Delta_m(s) = e^{-Ts} - 1 \qquad (27)$$

Figure 12 illustrates that with the choice of $\omega_c = 200 rad/s$, the system is stable as blue line is below the red one with no intersection.

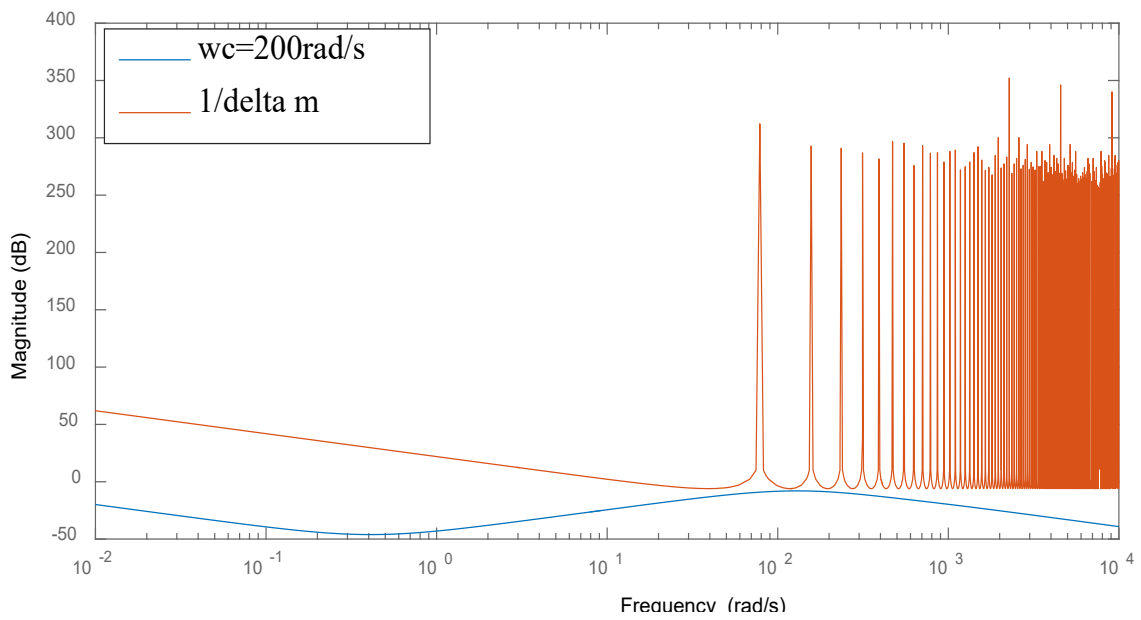

Figure 12. Magnitude of $\frac{L_n(j\omega)}{1+L_n(j\omega)}$ and $\frac{1}{\Delta_m}$ for stability of robustness

## VII Double Disturbance Observer

In order to deal with disturbance rejection and time delay simultaneously, DDOB compensated control system was used and its structure was depicted in figure 13. The lower block has the same structure as the CDOB and the upper block is a disturbance observer for disturbance rejection.

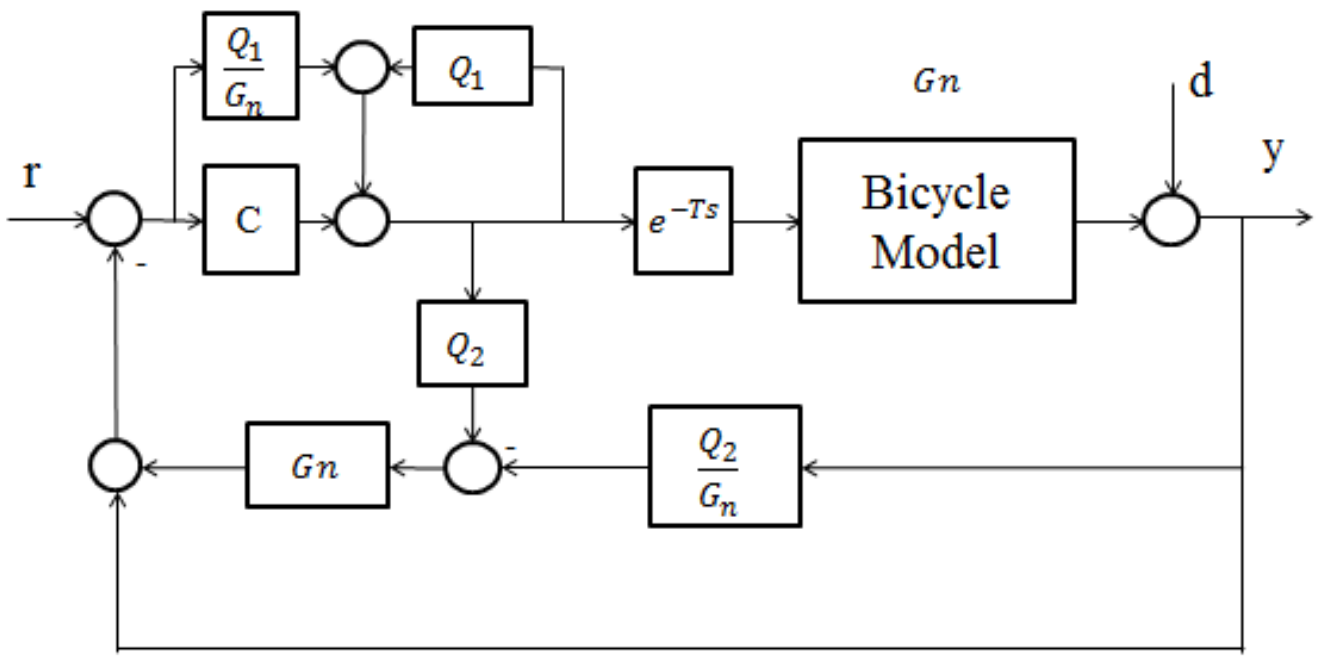

Figure 13. Double disturbance observer compensated control system

## VIII. Simulation Studies

Simulations are performed to check the performance enhancement in the autonomous vehicle path following control with proposed DOB and CDOB structure. The desired path to be followed is an elliptical route as shown in Figure 14 and the curvature of the path is depicted in Figure 15. Figure 16 to Figure 20 compares the path following errors of robust PD feedback controller system with and without disturbance observer compensation. For uncertain parameters, Figure 16 to Figure 19 takes the four corners of parametric uncertainty box into account. In Figure 20, external disturbance is added into the system due to road curvature input $\rho_{ref}$. It can be seen that with DOB added into the control system, the path following error decreases obviously as shown in Figure 16-20, which verify that DOB effectively deals with model regulation and disturbance rejection. Comparison about root-mean-square (RMS) errors of feedback control with DOB and feedback control only tabulated in Table II also illustrates the smaller errors in the presence of disturbance observer.

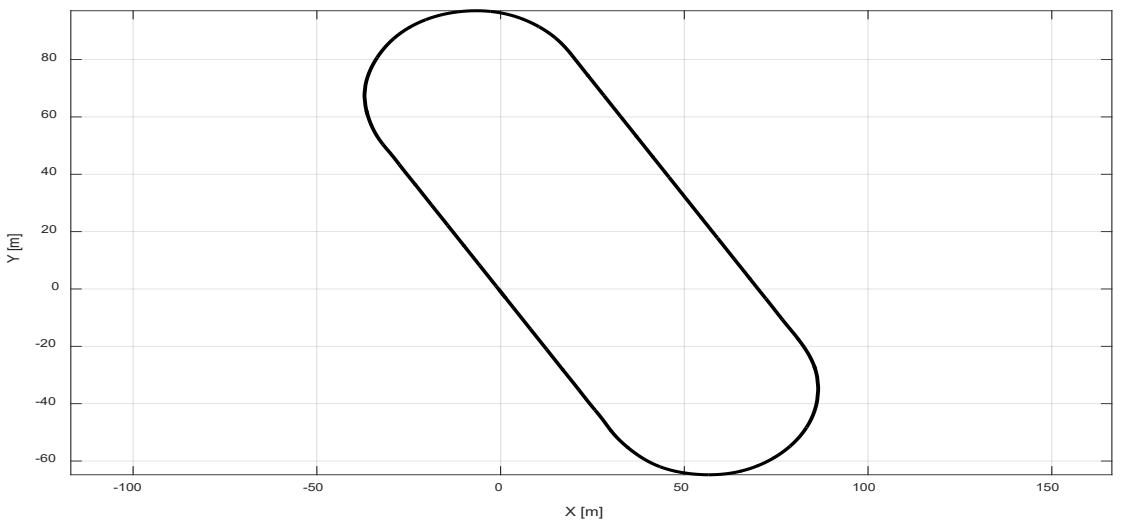

Figure 14. Desired path used in the simulation

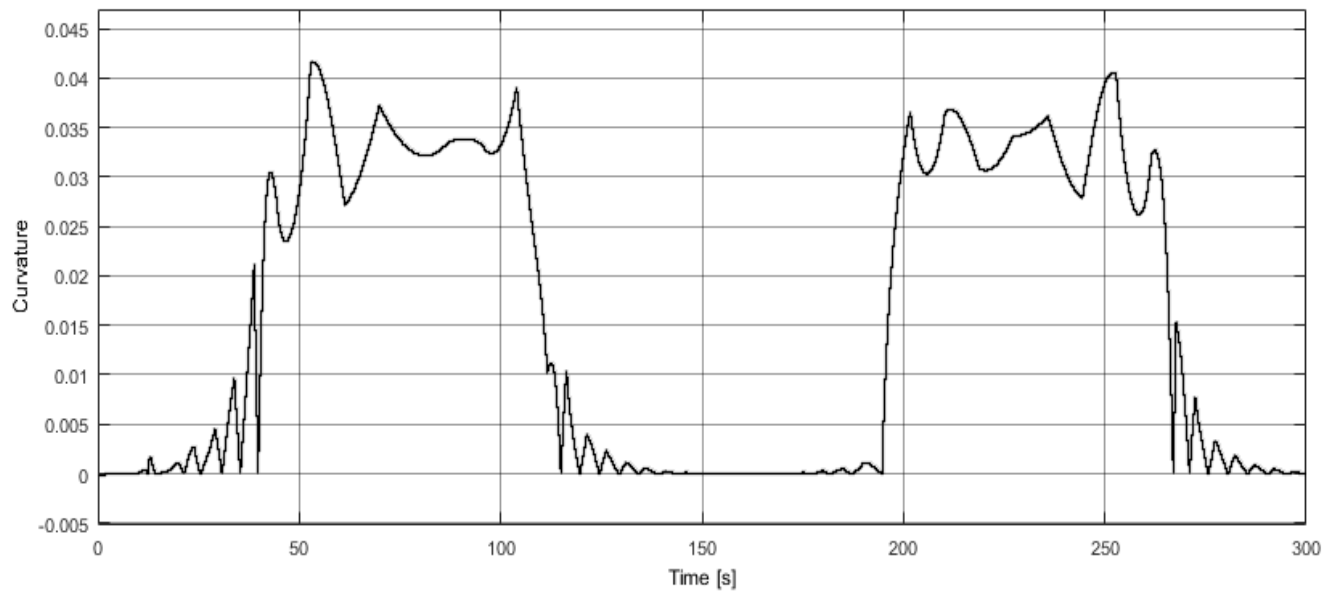

Figure 15. Curvature of the desired path

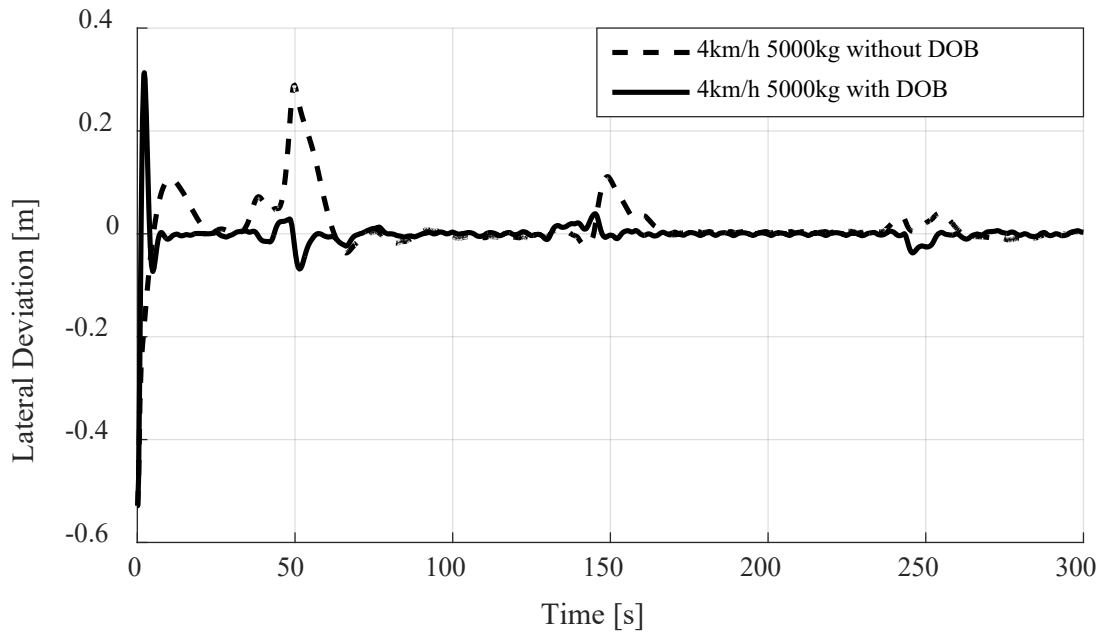

Figure 16. Lateral deviation with and without DOB at corner *a* for model uncertainty

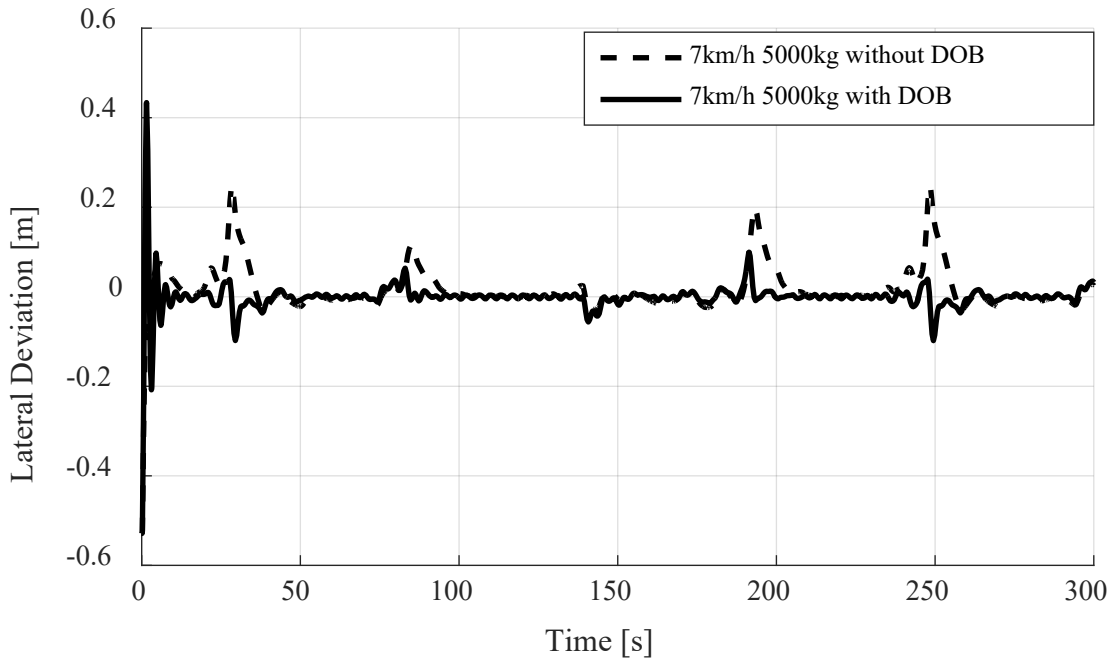

Figure 17. Lateral deviation with and without DOB at corner *b* for model uncertainty

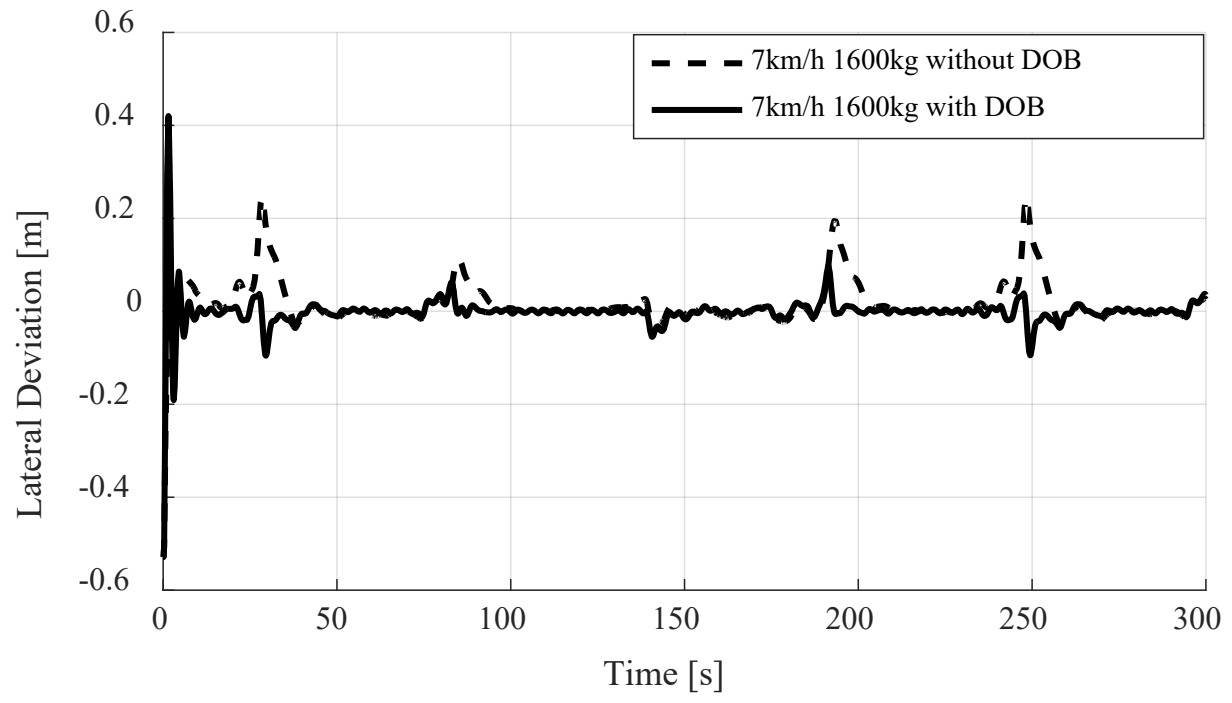

Figure 18. Lateral deviation with and without DOB at corner *c* for model uncertainty

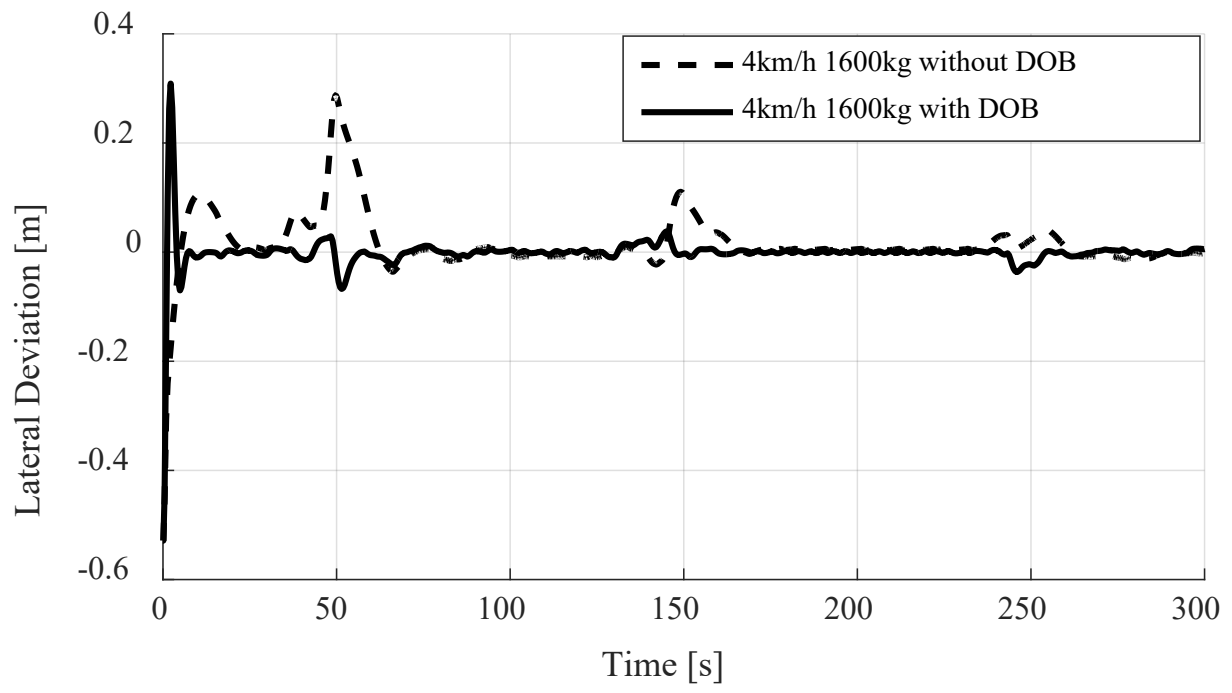

Figure 19. Lateral deviation with and without DOB at corner *d* for model uncertainty

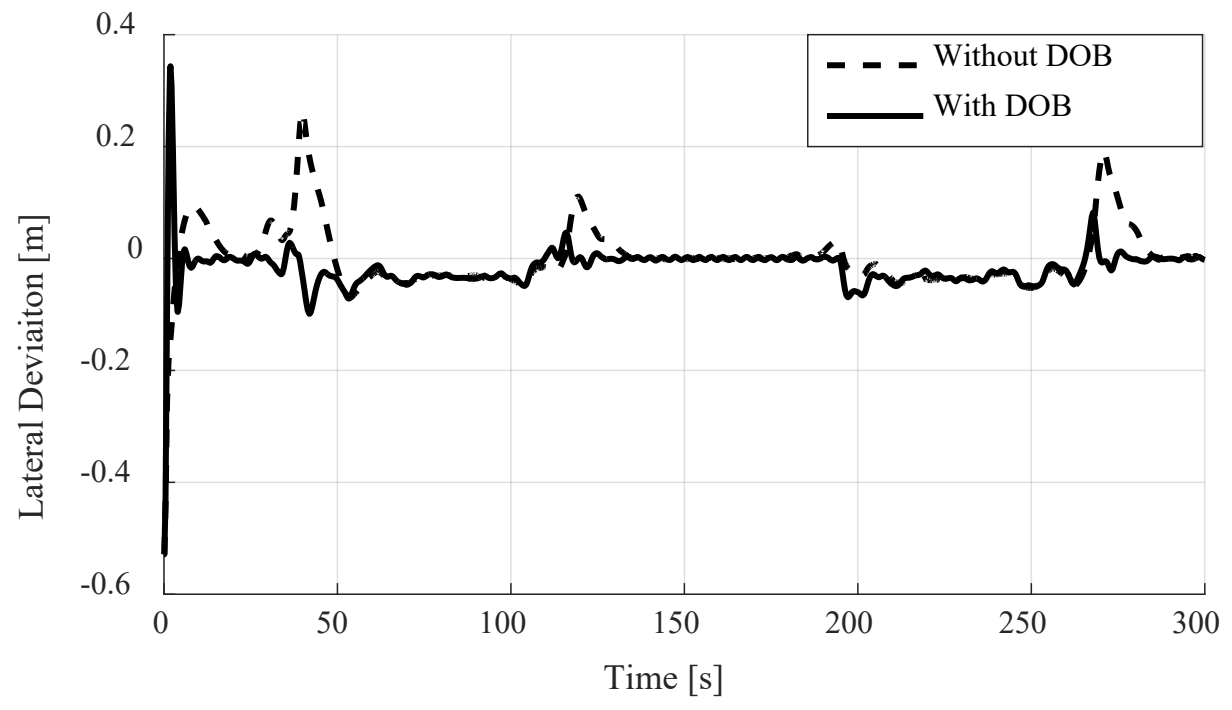

Figure 20. Lateral deviation with and without DOB for disturbance input

Table II Comparison of RMS tracking errors between PD and PD with DOB

| | 4km/h 1600kg | 4km/h 5000kg | 7km/h 1600kg | 7km/h 5000kg |
|---|---|---|---|---|
| PD | 0.0580 | 0.0581 | 0.0523 | 0.0526 |
| PD+DOB | 0.0320 | 0.0336 | 0.0359 | 0.0370 |

Figure 21 compares the lateral deviation of system with and without communication disturbance observer compensation by considering 0.08 sec CAN bus delay for steering actuation. It shows that CDOB compensates the time delay effect in the closed loop system and has reduced errors. From Figure 22, we can see that CDOB compensated control system has smaller steering angle compared with PD only controlled system. These results show a better path following performance of CDOB compensation control system.

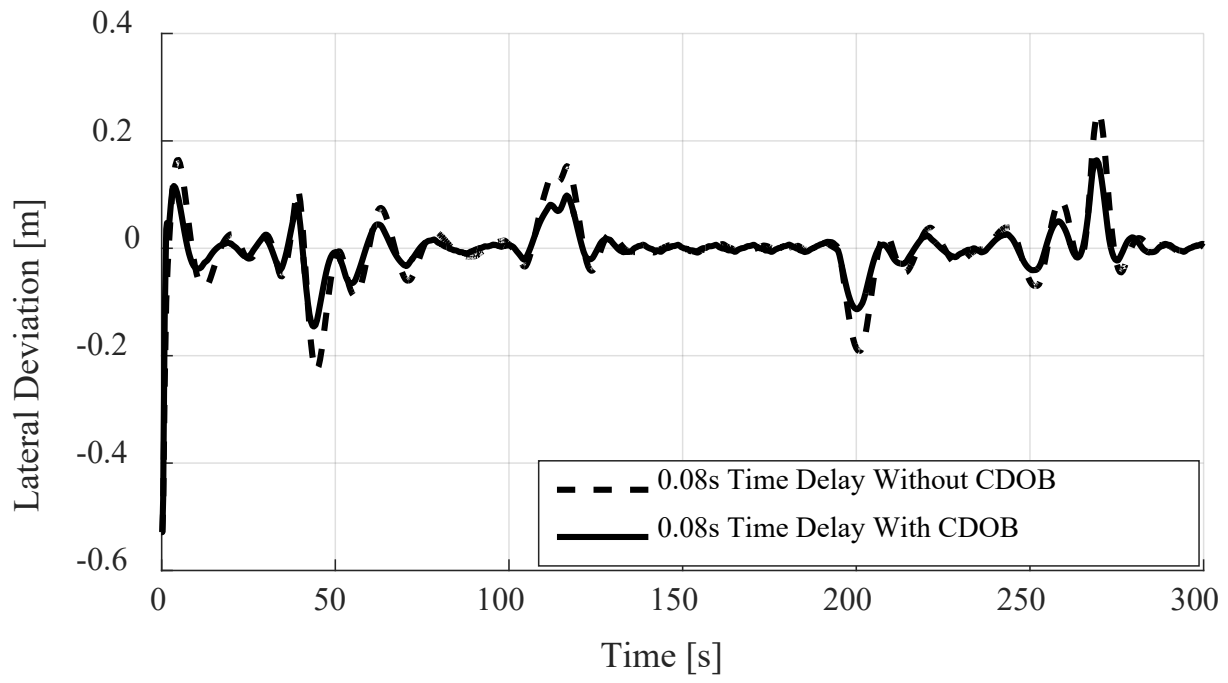

Figure 21. Lateral deviation with and without CDOB for time delay

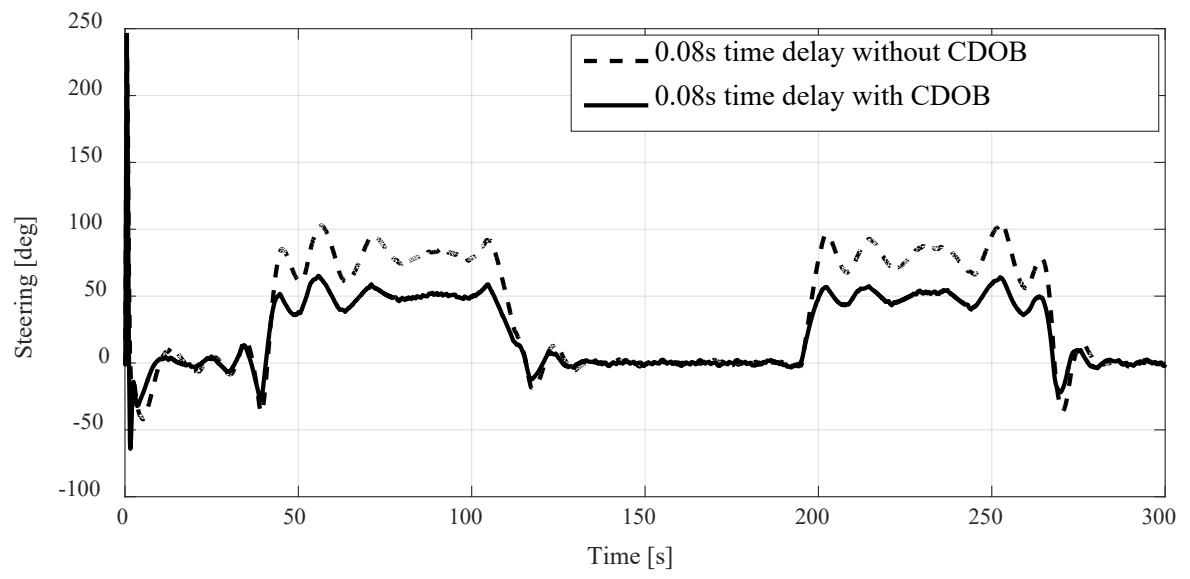

Figure 22. Steering angle with and without CDOB for time delay

Figure 23 and Figure 24 compare the lateral deviation and steering angle and speed of CDOB and DDOB compensated system when both 0.08sec time delay and disturbance input exist in the system simultaneously. It can be seen that both systems have similar steering angle and DDOB works better than CDOB with reduced path following errors.

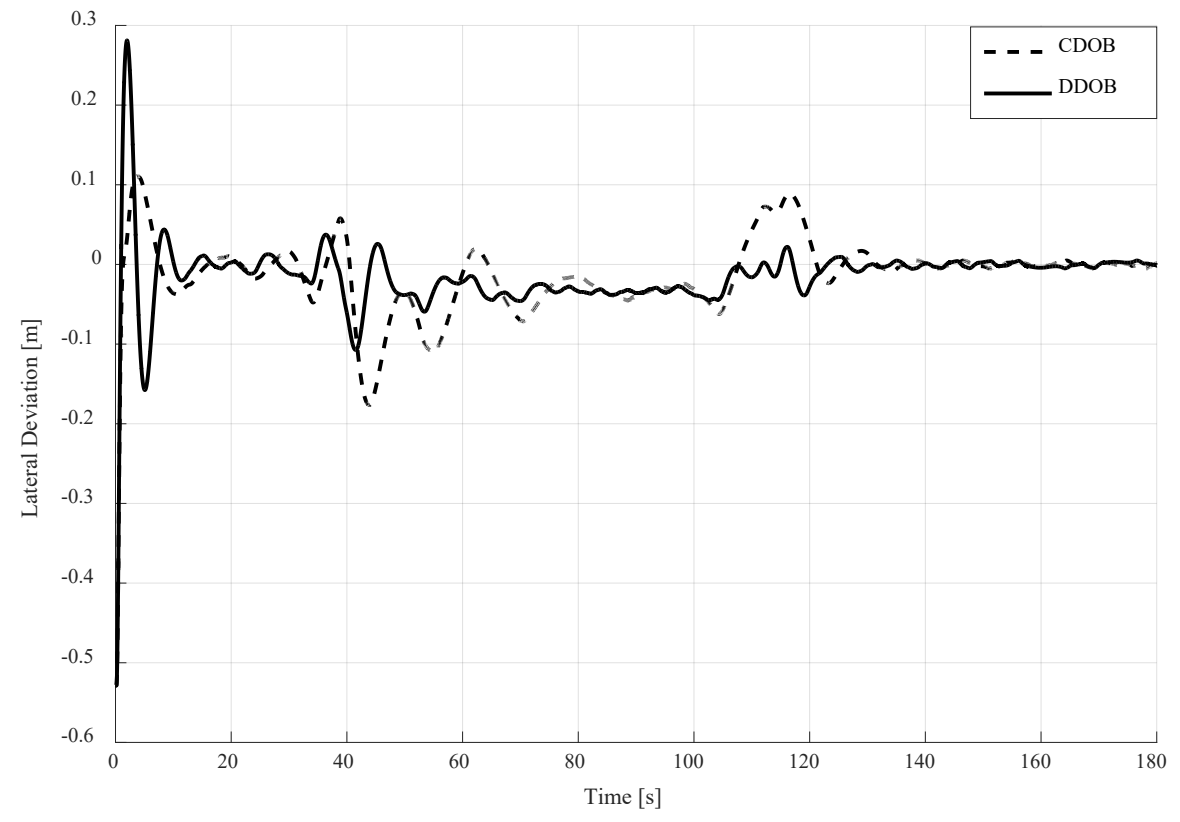

Figure 23 lateral deviation of CDOB and DDOB compensated system

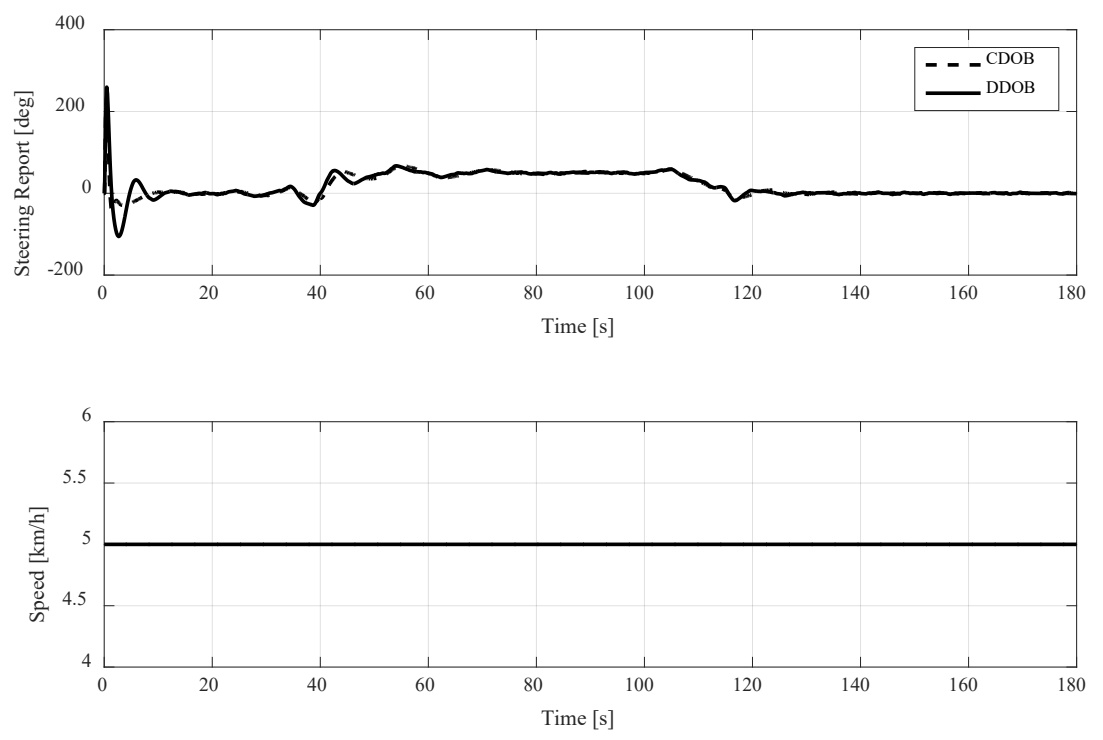

Figure 24 Steering angle and speed of CDOB and DDOB compensated system

# IX. Conclusion and future work

In this paper, the disturbance observer was applied to deal with model uncertainty and external disturbance and communication disturbance observer was used to handle CAN bus delay in order to realize performance enhancement of autonomous vehicle path following control. Also, double disturbance observer was applied in the vehicle path following control system to achieve model regulation, disturbance rejection and time delay simultaneously. Robust PD controller was designed based on the nominal model and *Q* filter design was presented. Robust stability of DOB and CDOB was studied analytically and verified. Simulation results were given to evaluate the vehicle path following performance and verify the proposed control algorithm.

In the future work, varying time delay will be studied with CDOB compensated system. More model-in-the-loop and hardware-in-the-loop (HiL) simulations will be performed to further test the designed DOB , CDOB and DDOB systems. Future work can also focus on and treat integrating or combining some of the approaches in this paper with other control, AV, CV, ADAS, automotive control and other topics like those in references [18-88].

## Acknowledgments

This paper is based upon work supported by the National Science Foundation under Grant No.:1640308 for the NIST GCTC Smart City EAGER project UNIFY titled: Unified and Scalable Architecture for Low Speed Automated Shuttle Deployment in a Smart City, by the U.S. Department of Transportation Mobility 21: National University Transportation Center for Improving Mobility (CMU) sub-project titled: SmartShuttle: Model Based Design and Evaluation of Automated On-Demand Shuttles for Solving the First-Mile and Last-Mile Problem in a Smart City.

## Definitions/Abbreviations

| | |
|---|---|
| **AV** | Autonomous Vehicle |
| **HiL** | Hardware-in-the-Loop |